\documentclass[aps,pra,reprint,showpacs,superscriptaddress]{revtex4-1}

\usepackage{CJK}
\usepackage{etoolbox}
\usepackage{graphicx,amssymb,bm,amsfonts,amsmath,graphics,color,epstopdf}
\usepackage[bookmarks=false,pdfstartview=FitH,colorlinks=true,citecolor=blue,linkcolor=blue]{hyperref}
\newcommand {\bscco}{Bi$_2$Sr$_2$CaCu$_2$O$_{8+\delta}$}
\newcommand {\uJcm}{$\mu$J/cm$^2$}
\newcommand {\refeq}[1]{(\ref{#1})}

\AtBeginEnvironment{align}{\setcounter{subeqn}{0}}
\newcounter{subeqn} \renewcommand{\thesubeqn}{\theequation\alph{subeqn}}%
\newcommand{\subeqn}{%
  \refstepcounter{subeqn}
  \tag{\thesubeqn}
}

\begin{document}
\begin{CJK*}{GBK}{}


\title{Nonequilibrium Electron Dynamics in a Solid with a Changing Nodal Excitation Gap}

\author{Christopher L.\ Smallwood}
\email[Email: ]{chris.smallwood@colorado.edu}
\affiliation{Materials Sciences Division, Lawrence Berkeley National Laboratory, Berkeley, California 94720, USA}
\affiliation{JILA, University of Colorado \& National Institute of Standards and Technology, Boulder, Colorado 80309, USA}
\author{Tristan L.\ Miller}
\affiliation{Materials Sciences Division, Lawrence Berkeley National Laboratory, Berkeley, California 94720, USA}
\affiliation{Department of Physics, University of California, Berkeley, California 94720, USA}
\author{Wentao Zhang}
\altaffiliation[Present address: ]{Shanghai Jiao Tong University, Shanghai 200240, China}
\affiliation{Materials Sciences Division, Lawrence Berkeley National Laboratory, Berkeley, California 94720, USA}
\affiliation{Department of Physics, University of California, Berkeley, California 94720, USA}
\author{Robert A.\ Kaindl}
\affiliation{Materials Sciences Division, Lawrence Berkeley National Laboratory, Berkeley, California 94720, USA}
\author{Alessandra Lanzara}
\email[Email: ]{alanzara@lbl.gov}
\affiliation{Materials Sciences Division, Lawrence Berkeley National Laboratory, Berkeley, California 94720, USA}
\affiliation{Department of Physics, University of California, Berkeley, California 94720, USA}
\date {\today}

\begin{abstract}
We develop a computationally inexpensive model to examine the dynamics of boson-assisted electron relaxation in solids, studying nonequilibrium dynamics in a metal, in a nodal superconductor with a stationary density of states, and in a nodal superconductor where the gap dynamically opens. In the metallic system, the electron population resembles a thermal population at all times, but the presence of even a fixed nodal gap both invalidates a purely thermal treatment and sharply curtails relaxation rates. For a gap that is allowed to open as electron relaxation proceeds, effects are even more pronounced, and gap dynamics become coupled to the dynamics of the electron population. Comparisons to experiments reveal that phase-space restrictions in the presence of a gap are likely to play a significant role in the widespread observation of coexisting femtosecond and picosecond dynamics in the cuprate high-temperature superconductors.
\end{abstract}


\maketitle
\end{CJK*} 


%

\section{Introduction}

In the science of quantum materials, advances in ultrafast spectroscopy are facilitating new ways of extracting information related to equilibrium states of matter \cite{Gedik03,Perfetti07,Sobota12,Smallwood12,Zhang14}, and are in cases enabling the generation of metastable phases that had not previously existed \cite{Fausti11,Wang13,Stojchevska14}. To keep pace with these developments, there is an increasing need to develop nonequilibrium theories of electron dynamics in solids. This task is hampered by the fact that once a system is boosted out of equilibrium, standard thermodynamic quantities---including temperature, chemical potential, and specific heat---do not exist.

Nevertheless, a number of theoretical tools have emerged to examine nonequilibrium electron dynamics \cite{Kaganov57,Anisimov74,Allen87,Sun94,Groeneveld95,Bejan97,Gusev98,Lugovskoy99,Knorren00,DelFatti00,Rethfeld02,Kabanov08,Mueller13,Sobota14,Yang15,Rothwarf67,Owen72,Schmid74,Parker75,Kaplan76,Howell04,Axt04,Kira06,Papenkort07,Unterhinninghofen08,Freericks09,Sentef13,Moritz13,Kemper14,Kemper15,Peronaci15,Maghrebi16}, in recent years perhaps most prominently including calculations based on a Keldysh contour approach \cite{Freericks09,Sentef13,Moritz13,Kemper14,Kemper15,Peronaci15,Maghrebi16}. These models are beginning to be able to successfully model momentum- and energy-dependent band structure effects following an ultrafast femtosecond pulse in both metals and superconductors. For example, this approach has been used to demonstrate a suggestive relationship between nonequilibrium timescales and the imaginary part of the equilibrium self-energy \cite{Sentef13}, as well as to theoretically investigate Higgs mode signatures in a superconductor \cite{Kemper15}. 

Though Keldysh contour calculations are unparalleled in their sophistication and theoretical rigor, to date they are also computationally expensive, requiring supercomputer capabilities in order to be able to obtain useful results. For this reason, experimental efforts to examine electron dynamics in solid-state systems often rely on simpler models \cite{Kaganov57,Anisimov74,Allen87,Sun94,Groeneveld95,Bejan97,Gusev98,Lugovskoy99,Knorren00,DelFatti00,Rethfeld02,Kabanov08,Mueller13,Sobota14,Yang15,Rothwarf67,Owen72,Schmid74,Parker75,Kaplan76,Howell04}, in many cases postulating the existence of nonequilibrium temperature and/or chemical potential parameters, or integrating away all energy and momentum dependence so that dynamics can be described in terms of a single parameter for electron density.

In the present work, we study boson-assisted quasiparticle relaxation in solids using an intermediate Fermi's golden rule approach \cite{Grimvall,Bauer15}, which is computationally more tractable than Keldysh contour models, yet still more sophisticated than the most popular phenomenological models. The model is energy-resolved, and can also incorporate an electronic excitation gap with a magnitude that changes as a function of time. This is particularly relevant for materials that display a photoexcitation-sensitive band gap in the electronic spectra, such as superconductors \cite{Beck11,Smallwood12,Smallwood14} and charge density wave systems \cite{Perfetti06,Schmitt08,Rohwer11}. We show that many of the model's characteristic predictions bear a striking resemblance to experimental findings of quasiparticle relaxation in the cuprate superconductor \bscco\ (Bi2212), observed via time-resolved ARPES and pump-probe spectroscopy. In particular, we demonstrate that the widespread observation of two-component relaxation dynamics in cuprates at high fluences is likely to be heavily influenced by the presence and dynamics of phase-space restrictions. Beyond this, our hope is that the model will be useful to experimentalists and theorists alike as a tool in understanding quasiparticle relaxation in other types of materials exhibiting band structure with Dirac nodes, such as graphene or topological insulators.

\section{Model}

A simulation is constructed to capture the effects of a standard pump-probe experiment, in which a solid-state system is driven out of equilibrium by an ultrafast optical pump pulse, and is probed by a second optical pulse at a later point in time that reads out the nonequilibrium quasiparticle population in the form of a transmissivity, reflectivity, or photoemission signal. We assume that the probe pulse arrives sufficiently later than the pump pulse so that the detailed time dependences of the pump pulse's electric and magnetic fields do not need to be taken into account in determining time-dependent quasiparticle evolution.

Within this framework, we assume that the energy-dependent population of electronic quasiparticles $P(\omega,t)$ can be written in terms of the product
\begin{equation}
P(\omega,t) = D(\omega,t)f_e(\omega,t) \label{eq1}
\end{equation}
where $D(\omega,t)$ is a time-dependent density of states, and $f_e(\omega,t)$ corresponds to a fermionic distribution function with a restricted range such that $0 \leq f_e(\omega,t) \leq 1$. For the sake of simplicity, we ignore momentum dependence. The evolution of the quasiparticle population $P(\omega,t)$ at subsequent times is determined, as discussed below in greater detail, through iterated applications of letting the distribution function $f_e(\omega,t)$ evolve according to collision integrals, forcing the density of states $D(\omega,t)$ to change in response to the updated $f_e(\omega,t)$, and then readjusting $f_e(\omega,t)$ to accommodate the modified density of states $D(\omega,t)$ in a manner such that the integral of $P(\omega,t)$ with respect to energy is unaffected by the change in $D(\omega,t)$. 

In essence, our model follows the spirit of a Boltzmann approach \cite{Kaganov57,Anisimov74,Allen87,Sun94,Groeneveld95,Bejan97,Gusev98,Lugovskoy99,Knorren00,DelFatti00,Rethfeld02,Kabanov08,Mueller13,Sobota14,Yang15}. We place no restrictions on the functional form of $f_e(\omega,t)$ beyond its initial condition, however, and this importantly distinguishes the present work from $N$-temperature models that are more widely discussed in the literature \cite{Kaganov57,Anisimov74,Allen87,Brorson90,Bovensiepen07}, in which the Boltzmann equation has been reduced further by forcing the nonequilibrium electron distribution to take on a thermal profile at every step in time. In fact, we will show here that in some of the most relevant cases, $f_e(\omega,t)$ explicitly cannot be described in thermal terms. 

It is convenient to further divide out the density of states $D_0$ at the chemical potential in the metallic state from $D(\omega,t)$, such that $D(\omega,t) = D_0 D_1(\omega,t)$, which implicitly defines a unitless normalized density of states $D_1(\omega,t)$, and a unitless normalized population density $P_1 \equiv D_1(\omega,t) f_e(\omega,t)$. For a decay process dominated by boson absorption and emission, Fermi's golden rule dictates that the distribution function $f_e(\omega,t)$ should evolve with time according to
\begin{align}
&\frac{\partial f_e(\omega)}{\partial t} = -\frac{2 \pi}{\hbar} \int_0^\infty d\Omega \, \alpha^2 F(\Omega) \, \times \label{fermigolden} \\
\big\{& D_1(\omega-\Omega) f_e(\omega)[1- f_e(\omega-\Omega)][n(\Omega)+1] \subeqn \label{fermigolden:a}\\
-& D_1(\omega-\Omega) [1-f_e(\omega)]f_e(\omega-\Omega)n(\Omega) \subeqn \\
+& D_1(\omega+\Omega) f_e(\omega)[1- f_e(\omega+\Omega)]n(\Omega) \subeqn \\
-& D_1(\omega+\Omega) [1-f_e(\omega)] f_e(\omega+\Omega) [n(\Omega)+1]\big\}, \subeqn \label{fermigolden:d}
\end{align} 
where $\alpha^2 F(\Omega)$ is the Eliashberg coupling function, $n(\Omega) = n(\Omega,t)$ is an optionally time-dependent bosonic distribution function that reduces to the Bose-Einstein distribution function at equilibrium, and where the quantities $\omega$ and $\Omega$ carry units of energy \cite{Grimvall}. In taking the formalism from Eq.~(\ref{eq1}) to Eq.~(\ref{fermigolden}), $D_0$ has been absorbed into the definition of $\alpha^2F(\Omega)$.

\begin{figure}[tb]\centering\includegraphics[width=3.375in]{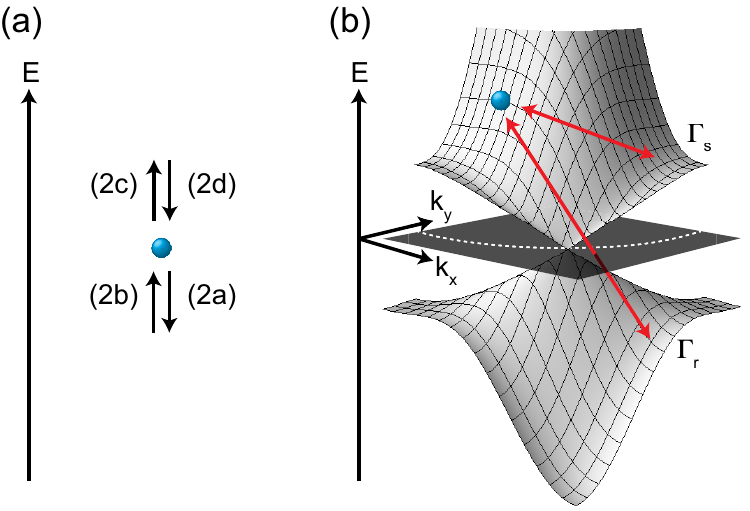}
\caption{\label{cartoon}Two ways of classifying electronic quasiparticle relaxation dynamics.
{\bf(a)} The total change in the quasiparticle distribution function $f_e(\omega,t)$ is given by the sum of four interactions under a Fermi's golden rule analysis, defined by terms \refeq{fermigolden:a}--\refeq{fermigolden:d} in Eq.~\refeq{fermigolden}.
{\bf(b)} The interactions can also be grouped into a scattering rate $\Gamma_s$ and recombination rate $\Gamma_r$, defined according to Eqs.~\refeq{fermigolden2}--\refeq{recomb}.
}
\end{figure}

Eq.~\refeq{fermigolden} can be decomposed into four straightforward physical processes. As illustrated by Fig.~\ref{cartoon}(a), the bracketed terms \refeq{fermigolden:a}--\refeq{fermigolden:d} respectively correspond to quasiparticle transitions via boson emission away from a state at energy $E=\omega$ to lower energies, transitions via boson absorption into the state at energy $E=\omega$ from lower energies, transitions via boson
absorption away from the state at energy $E=\omega$ to higher energies, and transitions via boson emission into the state at energy $E=\omega$ from higher energies.

Before proceeding, we note a few caveats to this model and parallels to related studies. First, the model is motivated by a desire to understand nonequilibrium dynamics in cuprate superconductors, which means that a more complete theoretical treatment should include coherence factors $C(\omega,\omega\pm\Omega,\Delta_k,\Delta_{k'})$ (see Refs.~\cite{Tinkham} and \cite{Hanaguri09}) before each of the terms in Eq.~\refeq{fermigolden}. Indeed, there is some evidence that coherence factors play a role in the temperature dependence of quasiparticle scattering rates for underdoped cuprates \cite{Hinton16}. We have ignored such effects in this work because they average to unity in a momentum-integrated picture of a $d$-wave superconductor with isotropic scattering. Second, in addition to electron-boson interactions, quasiparticle relaxation may be governed by electron-electron and electron-impurity interactions. Although not treated in the present work, these interactions may be relevant, particularly at short times. Potential effects have been considered within the context of cuprate superconductors by the authors of Refs.~\cite{Gedik04,Howell04,Perfetti07,Yang15}. Finally, being an energy-resolved yet momentum-integrated model motivated by superconductivity, we note that Eq.~\refeq{fermigolden} shares characteristics with previous work by Kaplan {\it et al.}\ \cite{Kaplan76}, who used an energy-dependent model to study quasiparticle dynamics in $s$-wave superconductors. The most substantive difference between the two works is that Kaplan {\it et al.}\ consider the dynamics of quasiparticles in a near-equilibrium system, whereas the present work is motivated by dynamics far from equilibrium. Hence, Kaplan {\it et al.}\ do not need to treat the possibility of a dynamically changing gap. Beyond this, it should perhaps be noted that the lifetimes calculated by Kaplan {\it et al.}\ correspond to the imaginary part of the electronic self-energy, whereas those in the present work are connected to $\partial f_e(\omega) / \partial t$, which is a qualitatively different parameter (see Ref.~\cite{Grimvall} and Appendix \ref{sec:selfenergy} for further discussion).

Returning to the model, it is in cases useful to regroup Eq.~\refeq{fermigolden} according to
\begin{equation}
\frac{\partial f_e(\omega)}{\partial t} = -(  \Gamma_s +  \Gamma_r), \label{fermigolden2}
\end{equation}
defining a scattering rate $\Gamma_s$ as
\begin{align}
\Gamma_s(\omega) &\equiv \frac{2 \pi}{\hbar} \int_0^\omega d\Omega \, \alpha^2 F(\Omega) \,  D_1(\omega-\Omega) \times \label{scatter}\\
&\qquad \quad \{ f_e(\omega)[1- f_e(\omega-\Omega)][n(\Omega)+1] \nonumber \\
&\qquad \quad - [1-f_e(\omega)]f_e(\omega-\Omega)n(\Omega)\} \nonumber \\
&+\frac{2 \pi}{\hbar} \int_0^\infty d\Omega \, \alpha^2 F(\Omega) \, D_1(\omega+\Omega) \times \nonumber \\
&\qquad \quad \{  f_e(\omega)[1- f_e(\omega+\Omega)]n(\Omega) \nonumber \\
&\qquad \quad -[1-f_e(\omega)] f_e(\omega+\Omega) [n(\Omega)+1]\}, \nonumber \\
\intertext{and a recombination rate $\Gamma_r$ as}
\Gamma_r(\omega) &\equiv \frac{2 \pi}{\hbar} \int_\omega^\infty d\Omega \, \alpha^2 F(\Omega) \, D_1(\omega-\Omega) \times \label{recomb}\\
&\qquad \quad \{ f_e(\omega)[1- f_e(\omega-\Omega)][n(\Omega)+1] \nonumber \\
&\qquad \quad -[1-f_e(\omega)]f_e(\omega-\Omega)n(\Omega)\}. \nonumber
\end{align} 
In this way, processes in which quasiparticle number is conserved (described by $\Gamma_s$) have been explicitly separated from those involving pair-breaking or pair recombination (described by $\Gamma_r$). The distinction between the two types of scattering channels is depicted pictorially in Fig.~\ref{cartoon}(b).

\section{Decay rates}
\label{sec:decayrates}

\begin{figure}[tb]\centering\includegraphics[width=3.375in]{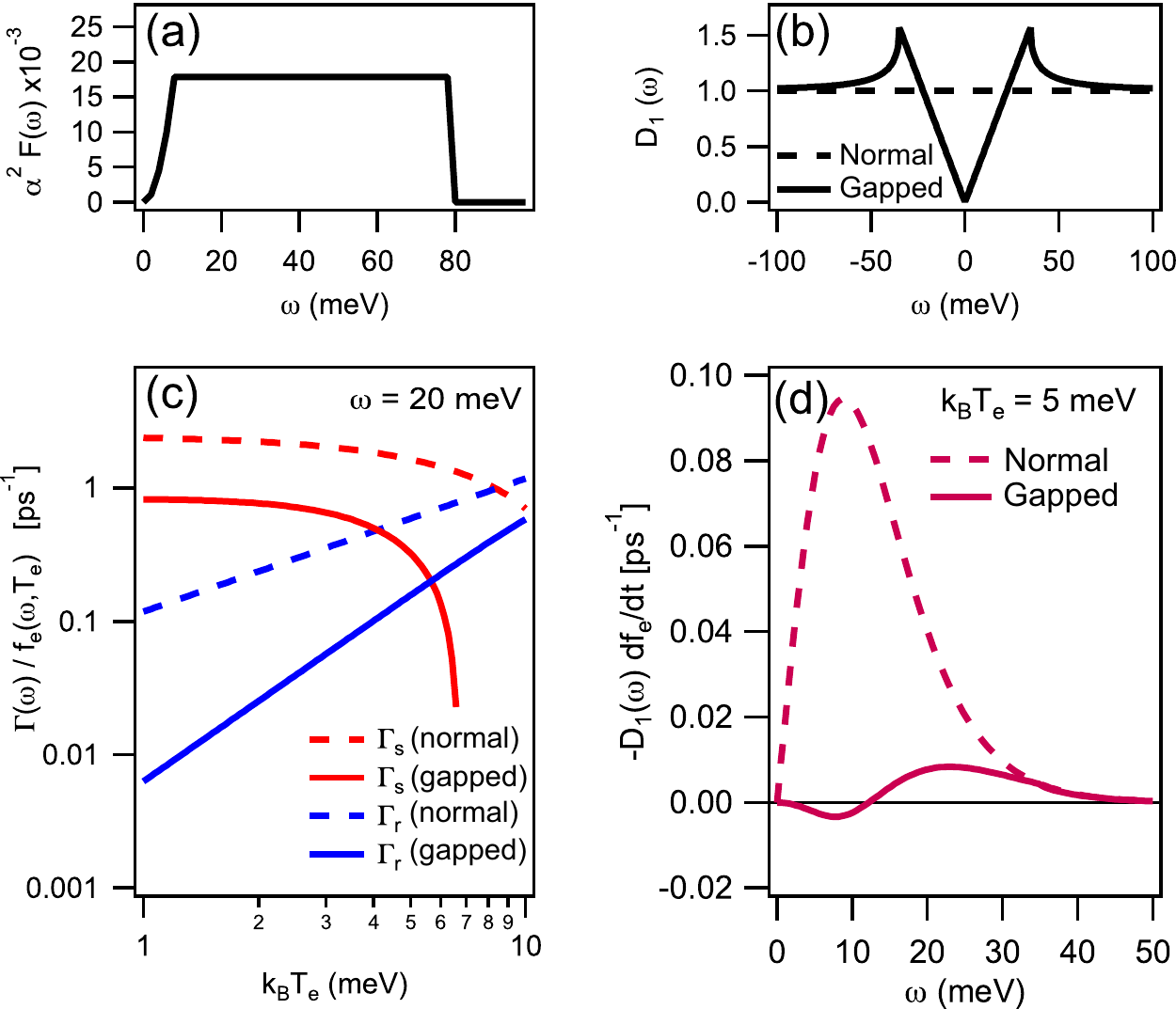}
\caption{\label{rates}Comparison between quasiparticle scattering and recombination rates for a metallic system and a nodal superconductor.
{\bf(a)} The Eliashberg coupling function is defined according to Eq.~\refeq{eliashberg} and normalized to result in a mass enhancement parameter $\lambda = 0.1$.
{\bf(b)} Normalized density of states $D_1(\omega)$, which is set to unity in the metallic system, and described by Eq.~\refeq{d1} in the gapped system.
{\bf(c)} Normalized scattering rate $\Gamma(\omega)/f_e(\omega,T_e)$ for $\omega = 20$ meV [where $\Gamma_s$ and $\Gamma_r$ are defined according to Eqs.~\refeq{scatter} and \refeq{recomb}] as a function of initial electronic temperature $k_BT_e$, assuming a lattice temperature of $k_BT_b = 0.1$ meV.
{\bf(d)} Total scattering rate $D_1 \partial f_e /\partial t$ for $k_BT_e$ = 5 meV as a function of quasiparticle energy $\omega$, assuming a lattice temperature of $k_BT_b = 0.1$ meV.
}
\end{figure}

To analyze the impact of the presence or absence of a band gap on quasiparticle relaxation rates more fully, we now adopt specific functional forms for $\alpha^2F(\Omega)$ and $D_1(\omega,t)$. We take the following form for $\alpha^2F(\Omega)$:
\begin{equation}
\alpha^2F(\Omega) = \left\{ 
\begin{array}{ll}
A \Omega^2 &\text{for } 0 \leq \Omega < \Omega_1 \\[5pt]
A \Omega_1^2 &\text{for } \Omega_1 \leq \Omega < \Omega_{max}
\end{array}
\right. , \label{eliashberg}
\end{equation}
where $\Omega_1 = 8$ meV and $\Omega_{max} = 80$ meV. The function is plotted in Fig.~\ref{rates}(a), and is consistent with the assumption of an energy-independent electron-boson coupling matrix element, as well as a relatively constant bosonic density of states $F(\Omega)$ that terminates at 80 meV. The latter assumption is consistent with measurements of the phonon density of states in Bi2212 by Renker, et al.\ \cite{Renker89}. The former assumption is chosen for its simplicity, although it may fail to capture some important effects, such as the dispersion kink, due to electron-phonon coupling, which is known to exist in almost all cuprates at 70 meV\@ \cite{Lanzara01}. The pre-factor $A$ is chosen to make the low-temperature limit of the mass enhancement parameter $\lambda$, which is connected to the Eliashberg coupling function according to the equation \cite{Grimvall}
\begin{equation}
\lambda = 2 \int_0^\infty \frac{\alpha^2 F(\Omega)}{\Omega} d\Omega,
\end{equation}
equal to 0.1\@.
We note that in choosing an Eliashberg function restricted to 80 meV, our analysis is most strongly motivated by the interactions between electrons and phonons, rather than between electrons and magnons or more exotic types of bosonic excitations that extend to higher energies. We are concerned in the present work with the dynamics of electrons at relatively low energies ($\omega<50$ meV) and long timescales ($t>100$ fs), for which the 80-meV cutoff is largely inconsequential.

To model quasiparticle recombination in the metallic state, we choose the featureless value $D_1(\omega,t) = 1$ [implicitly assuming a negligible metallic-state energy dependence to $D(\omega,t)$]. To model quasiparticle recombination in the gapped state, we choose a $d$-wave functional form
\begin{equation}
D_1(\omega,t) = \left\{ 
\begin{array}{ll}
\pi |\omega| / (2 \Delta_0) &\text{for } |\omega| \leq \Delta_0 \\[5pt]
(\omega/\Delta_0) \arcsin(\Delta_0/\omega) &\text{for } |\omega| > \Delta_0
\end{array}
\right. \label{d1}
\end{equation}
which is obtained by applying the standard Bardeen-Cooper-Schrieffer (BCS) relationship $E_k = \sqrt{\xi_k^2 + \Delta_k^2}$ \cite{Tinkham} to a cylindrical Fermi surface within a tetragonal Brillouin zone, where the gap parameter $\Delta_k$ depends in turn on Fermi surface angle $\phi_k$ according to $\Delta_k = 4 \Delta_0 (\phi_k - \pi/4) / \pi$ in the Brillouin zone's first quadrant, and is defined in the rest of the Brillouin zone according to the stipulation that $\Delta_\phi = -\Delta_{\phi+\pi/2}$. The parameter $\Delta_0$ is taken to be 35 meV\@. Plots of the metallic-state and gapped-state versions of $D_1(\omega,t)$ are depicted in Fig.~\ref{rates}(b).

Adopting these conventions, we now examine the effect that a nodal excitation gap has on quasiparticle relaxation rates. We set an equilibrium temperature $T = T_b = 0.1 \text{ meV} \approx 1$ K, we assume a large boson bath such that the nonequilibrium distribution function $n(\Omega,t)$ remains static in time and can be defined according to $n(\Omega) = 1/(e^{\Omega/k_BT_b}-1)$, and we imagine that the electronic distribution function $f_e(\omega,t)$ can be initially described by an elevated electronic temperature such that $f_e(\omega,t=0) = f(\omega,T_e) = 1/(e^{\omega/k_BT_e}+1)$. Figure \ref{rates}(c) shows the effect on relaxation rates at the initial time $t=0$ as a function of electronic temperature $T_e$, given a fixed quasiparticle energy of 20 meV. Figure \ref{rates}(d) shows relaxation rates at $t=0$ as a function of quasiparticle energy $\omega$, given a fixed electronic temperature of $k_BT_e = 5$ meV.

Figure~\ref{rates}(c) demonstrates for both the metallic and gapped states that the scattering rate $\Gamma_s$ remains essentially constant for low values of $k_B T_e$, and then begins to sharply decrease with increasing $k_B T_e$ as the characteristic energy $k_B T_e$ becomes more sizable. This can be understood as a consequence of Pauli blocking effects that suppress the ability of a quasiparticle to scatter into states at lower energies, and as a consequence of the fact that the rate for scattering into a state at energy $\omega$ from higher states increases with $k_BT_e$, owing to the greater number of occupied states at higher energies. Perhaps more interestingly, the recombination rates both for the metallic-state relaxation and gapped-state relaxation follow nearly perfect power laws as a function of $k_BT_e$, with $\Gamma_r \propto T_e$ in the metallic state and $\Gamma_r \propto T_e^2$ in the gapped state. In both cases, this power law dependence is a consequence of the second-order kinetics inherent in any quasiparticle recombination process, and can be described by the decoupled regime of the Rothwarf-Taylor model of quasiparticle recombination \cite{Rothwarf67}. It can in fact be shown that the Rothwarf-Taylor model is a special case of the present framework (see Appendix \ref{sec:rtmodel}).

The most important result of the decay rate analysis is the overall effect induced by a nodal gap. As shown in Fig.~\ref{rates}(c), the presence of a gap induces a sharp suppression of both scattering and recombination interactions, which is demonstrated by the fact that the solid blue and red lines lie significantly below their dashed counterparts. The effect is even more sharply pronounced in Fig.~\ref{rates}(d), where the change in the quasiparticle population $P_1(\omega,t) = D_1(\omega)f_e(\omega,t)$ is plotted as a function of energy. Whereas $-D_1(\omega) \partial f_e / \partial t$ is always positive in the metallic state and extends in cases to a rate of 0.1 ps$^{-1}$, the rates are sharply suppressed in the gapped state, and become in cases even negative.

\section{Quasiparticle evolution}
\label{sec:qpevolution}

Having analyzed energy-dependent quasiparticle decay rates at a fixed point in time, we proceed now with a more complete characterization of the quasiparticle population's temporal evolution. Results have been obtained numerically, by applying the Euler method to Eq.~\refeq{fermigolden} with $\alpha^2 F(\Omega)$ and $D_1(\omega)$ defined as in the previous section. As above, although the lattice distribution function $n(\Omega)$ could in principle be allowed to vary with time as it absorbs energy from electrons, we have for the sake of simplicity kept it fixed as a Bose-Einstein distribution function at a constant temperature $k_BT_b = 0.1$ meV. This is in agreement with a scenario where the bath of phonons in a material constitutes an essentially static reservoir that is affected only weakly by interactions with the electronic population.

\begin{figure}[tb]\centering\includegraphics[width=3.375in]{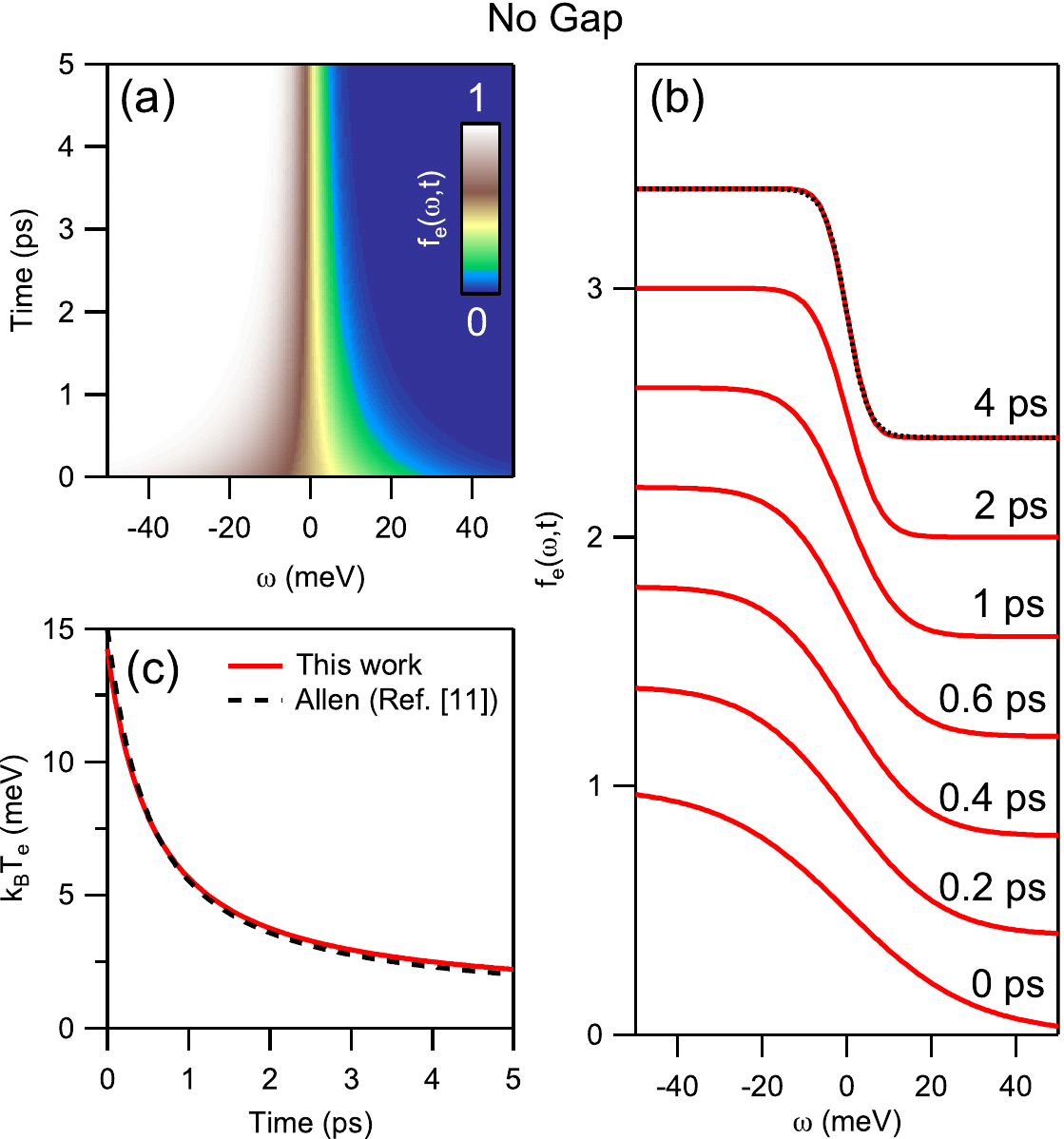}
\caption{\label{relaxmetal}Quasiparticle relaxation dynamics in a metal.
{\bf(a)} $f_e(\omega,t)$ vs.\ time and quasiparticle energy, given an equilibrium temperature $k_BT_b = 0.1$ meV, and an initially thermal quasiparticle population at a starting temperature of $k_BT_e = 15$ meV.
{\bf(b)} Selected energy distribution curves (EDCs) from (a), offset vertically for clarity. The dotted line overlapping the EDC at 4 ps is the result of fitting the data to a Fermi-Dirac distribution function.
{\bf(c)} Characteristic quasiparticle ``temperature'' [defined by Eq.~\refeq{telec}] vs.\ time. The dashed black line is a comparison to a two-temperature model derived by Allen \cite{Allen87}.
}
\end{figure}

We begin with the case of a metal. Figure~\ref{relaxmetal} shows the evolution of an electron population that is assumed to be initially thermal and at a temperature of $k_BT_e = 15$ meV, and which relaxes amid a constant density of states such that $D_1(\omega,t)=1$. The decision to use a Fermi-Dirac distribution for an initial condition was made largely for the sake of simplicity~\footnote{Other functional forms for the initial electronic distribution can be readily incorporated. See Supplemental Material available online at http://dx.doi.org/10.1103/PhysRevB.93.235107 for details.}, and is demonstrably false in the short-time limit of experimental pump-probe measurements of both metals and superconductors \cite{Fann92,Bovensiepen07,Perfetti07,Smallwood14}. Nevertheless, the approximation is reasonable beyond about 100 fs in many materials \cite{Bovensiepen07,Perfetti07,Smallwood14}, and is likely to be the result of increased electron-electron interactions in the limit of high energy and negligible phase-space restrictions. The impacts of these interactions become restricted at lower energies (typical Fermi liquid self-energies decrease with $\omega$ proportionally to $\omega^2$, for example \cite{Ashcroft}), and in the presence of a $d$-wave excitation gap \cite{Gedik04,Howell04}, which justifies the subsequent exclusion of electron-electron interactions from the model at longer times. Although it goes beyond the scope of the present work, one can obtain a more sophisticated treatment of electron dynamics by explicitly including the effect of a pump pulse and electron-electron interactions. Analyses such as this have been performed in metals by the authors of Refs.~\cite{Sun94,Groeneveld95,Bejan97,Gusev98,Lugovskoy99,Knorren00,DelFatti00,Rethfeld02,Kabanov08,Mueller13}.

As shown by Fig.~\ref{relaxmetal}(a), and perhaps more clearly by the selected horizontal slices depicted in Fig.~\ref{relaxmetal}(b), the electronic distribution function $f_e(\omega,t)$ resembles a Fermi-Dirac distribution function $f(\omega,T_e) = 1/(e^{\omega/k_BT}+1)$ at essentially all times, even though no effort has been made to enforce an explicit functional form for $f_e(\omega,t)$ apart from the initial condition.

Recognizing such a resemblance invites one to define a time-dependent electronic temperature 
\begin{equation}
k_BT_e(t) \equiv \frac{1}{\ln(2)} \int_0^\infty f_e(\omega,t) \, d\omega, \label{telec}
\end{equation}
which can be compared, for example, with an $N$-temperature model that forces the distribution function $f_e(\omega,t)$ to be exactly equal to $f[\omega,T_e(t)]$ at all points in time. Figure~\ref{relaxmetal}(c) shows a comparison between the results of the present model and a two-temperature model formulated by Allen \cite{Allen87} under identical initial conditions. The agreement is excellent. 

\begin{figure}[tb]\centering\includegraphics[width=3.375in]{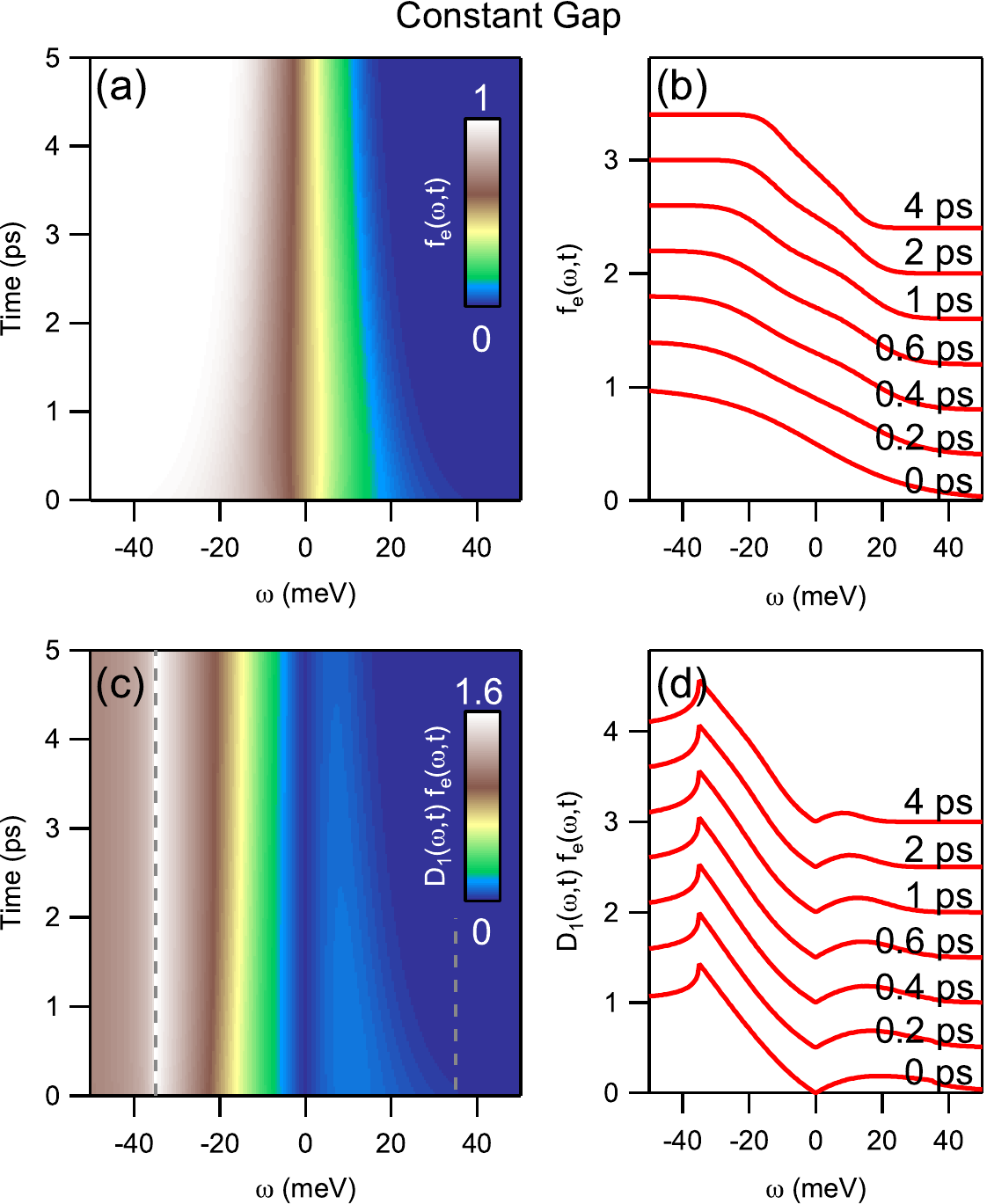}
\caption{\label{relaxd}Quasiparticle relaxation dynamics in a nodal superconductor with a fixed 35-meV gap.
{\bf(a)} $f_e(\omega,t)$ vs.\ time and quasiparticle energy, given an equilibrium temperature $k_BT_b = 0.1$ meV, and an initially thermal quasiparticle population at a starting temperature of $k_BT_e = 15$ meV.
{\bf(b)} Selected energy distribution curves (EDCs) from (a), offset vertically for clarity. 
{\bf(c)--(d)} Same as (a) and (b), where the quantity plotted is the energy-dependent quasiparticle density $D_1(\omega)f_e(\omega,t)$. The dashed gray lines in (c) correspond to the gap edge at $\pm \Delta_0$.
}
\end{figure}

\begin{figure}[tb]\centering\includegraphics[width=3.375in]{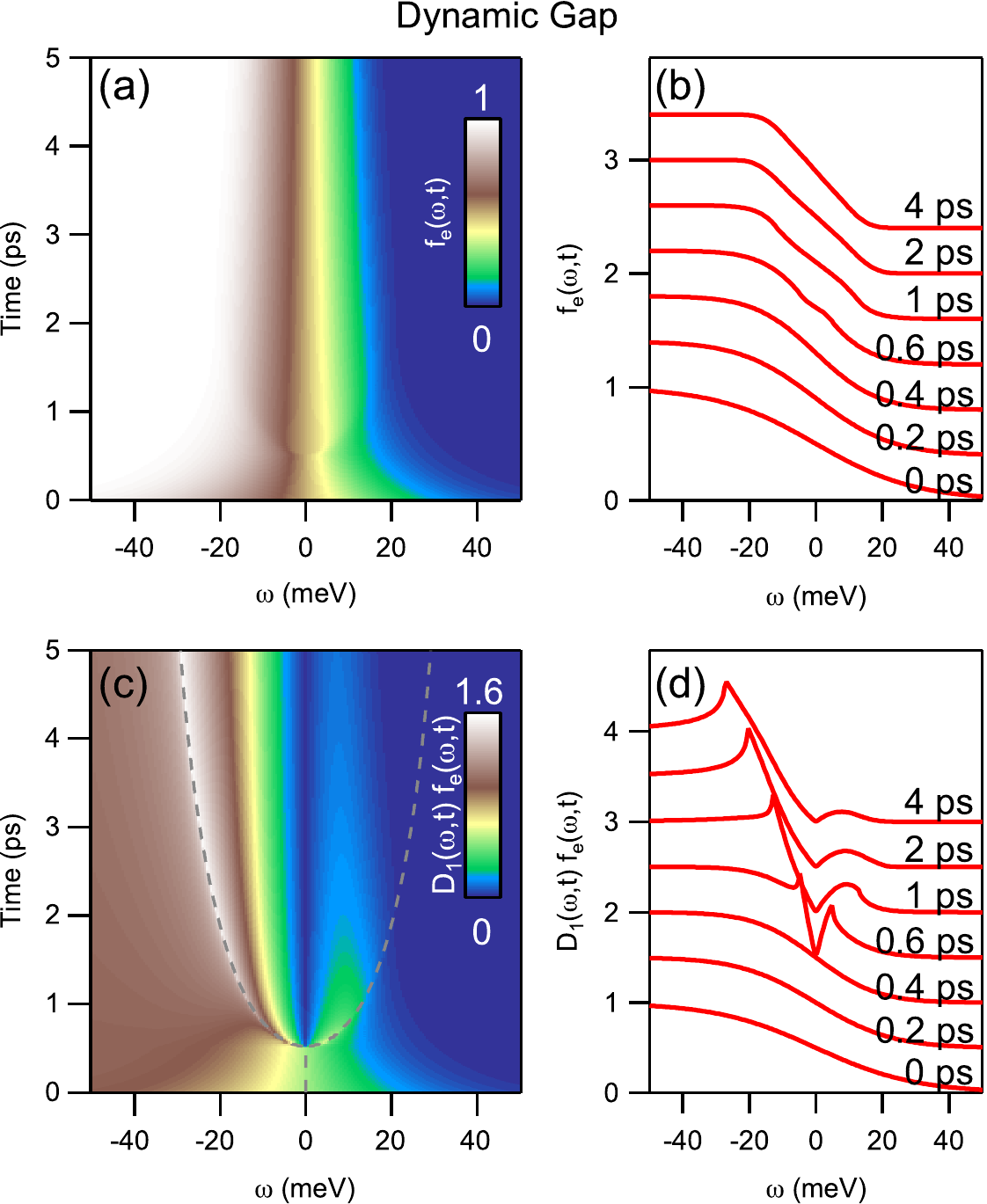}
\caption{\label{relaxd2}Quasiparticle relaxation dynamics in a nodal superconductor with a dynamic gap.
{\bf(a)} $f_e(\omega,t)$ vs.\ time and quasiparticle energy, given an equilibrium temperature $k_BT_b = 0.1$ meV, and an intially thermal quasiparticle population at a starting temperature of $k_BT_e = 15$ meV\@.
{\bf(b)} Selected energy distribution curves (EDCs) from (a), offset vertically for clarity.
{\bf(c)--(d)} Same as (a) and (b), where the quantity plotted is the energy-dependent quasiparticle density $D_1(\omega)f_e(\omega,t)$. The dashed gray lines in (c) correspond to the gap edge at $\pm \Delta(t)$.
}
\end{figure}

As shown in Figs.~\ref{relaxd} and \ref{relaxd2}, the situation becomes more interesting in the presence of a quasiparticle excitation gap. Figure~\ref{relaxd} shows the response of an initially thermal quasiparticle population at high temperature ($k_BT_e = 15$ meV) evolving toward equilibrium in the presence of a static $d$-wave gap of magnitude 35 meV [see Fig.~\ref{rates}(b)]. Panels (a) and (b) depict the evolution of the quasiparticle distribution function $f_e(\omega,t)$, while panels (c) and (d) depict the product $D_1(\omega) f_e(\omega,t)$, corresponding to the more complete energy-dependent quasiparticle population.

As shown, particularly in Fig.~\ref{relaxd}(b), even though $f_e(\omega,t)$ starts out (by construction) thermal at time $t=0$, it rapidly begins to develop additional structure in response to the gapped density of states. The deviation from thermal behavior is perhaps most evident at 0.6 and 1.0 ps, where the magnitude of the slope of $f_e(\omega,t)$ with respect to $\omega$ in the vicinity of $\omega=0$ is less than that at $|\omega| \approx 18$ meV\@. Such behavior is inconsistent with the shape of the Fermi-Dirac distribution function. Though less pronounced, deviations between $f_e(\omega,t)$ and a thermal distribution persist also at longer times.

As shown in Fig.~\ref{relaxd2}, the nonthermal dynamics that appear in the presence of a static gap are also prominent in a system that starts out metallic, but then develops a gap as the excitation density of quasiparticles decreases. Such is the case for an actual superconductor \cite{Smallwood14}. In order to simulate this dynamic gap, the electron population and density of states has been periodically adjusted in Fig.~\ref{relaxd2} between successive relaxation steps. The increase of the magnitude of the gap is phenomenologically determined by locking gap size to the effective electronic temperature parameter $T_e(\omega,t)$, which is defined as above in Eq.~\refeq{telec}, using the BCS gap equation \cite{Tinkham}, with a critical temperature of 7.8 meV\@. Conservation of quasiparticles between relaxation steps is achieved by adjusting the energy-dependent quasiparticle population with respect to changing gap size according to the equation
\begin{align}
P&_1(\omega,t_{i+1}) = \frac{\omega}{\Delta_{i+1}} \times  \label{openingdwave} \\[5 pt]
& \int_0^{\min(\omega,\Delta_{i+1}) } du \frac{ f_{e}\left(\sqrt{\omega^2 - [1-(\Delta_{i}/\Delta_{i+1})^2]u^2} ,t_i\right) }{\sqrt{\omega^2-u^2}}. \nonumber
\end{align}
This can be understood as a $d$-wave-gap extension of a condition for conserving the population of quasiparticles under a dynamic $s$-wave gap scenario that requires
\begin{equation}
f_e(\omega_{i+1},t_{i+1}) = f_e(\omega_{i},t_{i}),
\end{equation}
subject to the identity $\omega_{i+1}^2 - \Delta_{i+1}^2 = \omega_{i}^2 - \Delta_{i}^2$ (see Appendix \ref{sec:gapopen} for further details).

The dynamics captured by the simulation depicted in Fig.~\ref{relaxd2} can be largely understood as a hybrid between metallic relaxation dynamics and those in the presence of a static gap. As seen most easily in Fig.~\ref{relaxd2}(b), when the gap is closed, $f_e(\omega,t)$ relaxes thermally [see $f_e(\omega,t)$ for $t=0.2$ ps]. Once the gap begins to open ($t > 0.5$ ps), the distribution function deviates from a thermal distribution in a manner similar to that depicted in Fig.~\ref{relaxd}(b). 

Despite these similarities, the scenarios depicted in Figs.~\ref{relaxd} and \ref{relaxd2} are not perfectly analogous. An opening gap can actually reduce the energy-dependent rate of relaxation even more than would occur in the presence of a fixed gap if the gap is opening at a rate comparable to the rate of quasiparticle relaxation. The reason is that a dynamically opening gap actively lifts states from lower energy to higher energy as it opens. At energies inside the gap, where the density of states is decreasing with increasing time, this effect produces an increase in $f_e(\omega,t)$ with time, thereby effectively increasing the overall electronic temperature. Moreover, energy-dependent quasiparticle populations exhibit a brief increase in intensity as the gap edge transitions from below to above the energy in question.

\begin{figure}[tb]\centering\includegraphics[width=3.375in]{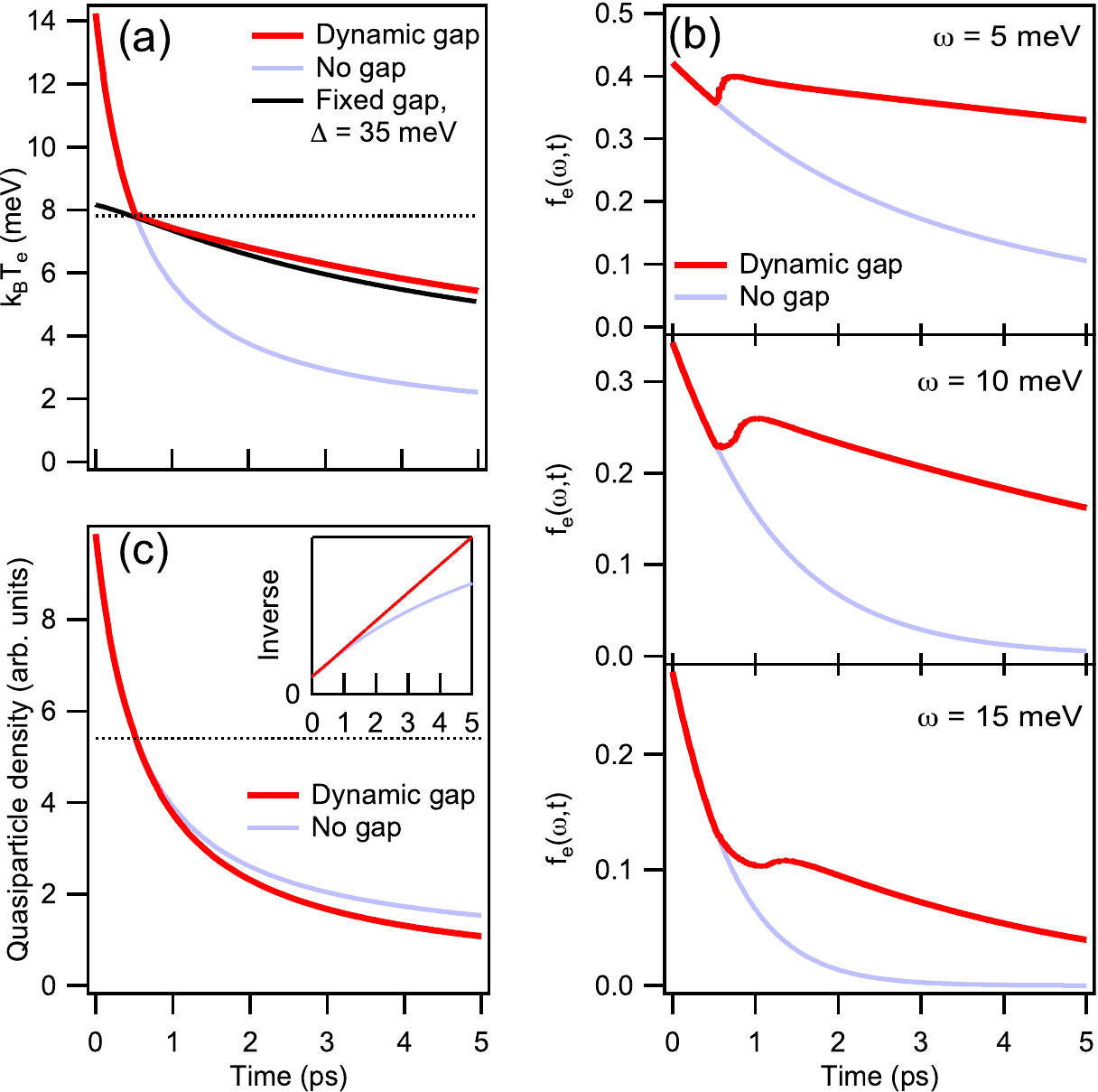}
\caption{\label{relaxtime}Comparison between the time dependence of quasiparticle relaxation dynamics in the presence of a dynamically opening nodal gap with those of a metal, given a quasiparticle population that is defined to be thermal at $t=0$, with an electronic temperature of $k_BT_e = 15$ meV\@. For the dynamically opening gap, $k_BT_c$ is defined as 7.8 meV. 
{\bf(a)} Effective electronic temperature $k_BT_e$ vs.\ time. For comparison, the evolution of $k_BT_e$ in the midst of a static 35 meV nodal gap is also plotted, scaled to an initial electronic temperature of 8.2 meV so that $k_BT_e$ crosses $k_BT_c$ at the same time that the effective electronic temperature crosses $k_BT_c$ in the dynamic gap scenario.
{\bf(b)} Energy-dependent quasiparticle population vs.\ time [vertical intensity profiles extracted from Figs.~\ref{relaxmetal}(a) and \ref{relaxd2}(c)].
{\bf(c)} Total quasiparticle population [integral of $D_1(\omega,t)f_e(\omega,t)$ from $\omega=0$ to $\omega=\infty$] vs.\ time.
}
\end{figure}

Predicted consequences can be observed in Fig.~\ref{relaxtime}, where the time-dependent effective temperature of the electronic state is directly compared among the no-gap, fixed-gap, and dynamic-gap cases. 
As shown in Fig.~\ref{relaxtime}(a), the no-gap and dynamic gap scenarios are identical at high excitation density, above the critical threshold for the gap to begin opening. By construction, the gap begins to open at $k_BT_e = 7.8$ meV, and a sharp reduction of the rate at which $k_BT_e$ decreases can be observed in the dynamic-gap scenario. A comparison of the red and black lines in Fig.~\ref{relaxtime}(a) reveals that the reduction of the decay rate for $k_BT_e$ under the dynamic gap scenario is in fact even more dramatic than it is under the static-gap scenario, as expected based on the arguments above. 

The influence of the dynamically opening gap can be further observed through an analysis of energy-dependent quasiparticle relaxation dynamics, depicted in Fig.~\ref{relaxtime}(b), which correspond to vertical quasiparticle intensity profiles extracted from Fig.~\ref{relaxd2}(c). Energy-dependent relaxation profiles extracted from Fig.~\ref{relaxmetal}(a) are also shown for comparison. The distinctive shift from rapid quasiparticle relaxation dynamics to more gradual relaxation dynamics can be observed by relative differences in the slope of the dynamic-gap relaxation curves before and after $t\approx 1$ ps. In addition, the energy-dependent quasiparticle dynamics experience an increase in population as the gap approaches the energy in question. Such effects of a dynamical gap, although perhaps surprising, are nevertheless replicated in more sophisticated nonequilibrium models. Similar quasiparticle signatures were seen, for example, by a recent Keldysh contour study of the effect of Higgs-mode gap oscillations on the quasiparticle spectrum \cite{Kemper15}.

It is worth noting that signatures of reduced scattering rates do not appear if, instead of characterizing an electronic temperature or energy-dependent population, one plots the entire energy-integrated quasiparticle population vs.\ time. Under this scenario, scattering (particle-conserving) events become irrelevant, and recombination (particle-annihilating) dynamics become the only relaxation events that contribute to the signal. Fig.~\ref{relaxtime}(c) shows that in both the metallic and dynamically-opening-gap scenarios, the quasiparticle population evolves smoothly across the threshold of $T_c$. For times longer than 0.5 ps, the metallic quasiparticle population actually decays more slowly than the gapped quasiparticle population. This can be understood as resulting from the fact that the Eliashberg coupling function drops to zero as $\omega \to 0$ [Fig.~\ref{rates}(a)], making low-energy quasiparticle recombination events, which are relevant to a metal, less probable than higher-energy quasiparticle recombination events, which are more relevant to a gapped superconductor. If the equilibrium temperature is sufficiently low, both metallic and gapped quasiparticle relaxation rates depend on quasiparticle density. In the gapped scenario, the relaxation dynamics follow a nearly perfect bimolecular recombination curve, as shown by the fact that the inverse of the total population density increases linearly with time [Fig.~\ref{relaxtime}(c) inset].

\section{Comparison to experiment}

Having described the model and its consequences in theoretical context, we now proceed with a comparison to experiments. We expect that the model will be relevant in helping to explain ultrafast relaxation dynamics in the high-temperature cuprate superconductors. It may be particularly useful in explaining a distinctive two-component quasiparticle relaxation dynamic that has been observed in superconducting Bi2212 using time-resolved ARPES \cite{Graf11,Cortes11,Piovera15,Smallwood15}, as well as similar two-component dynamics that have been observed in cuprates using all-optical pump-probe techniques \cite{Demsar99,Liu08a,Kusar08,Giannetti09,Coslovich11}.

\begin{figure}[tb]\centering\includegraphics[width=3.375in]{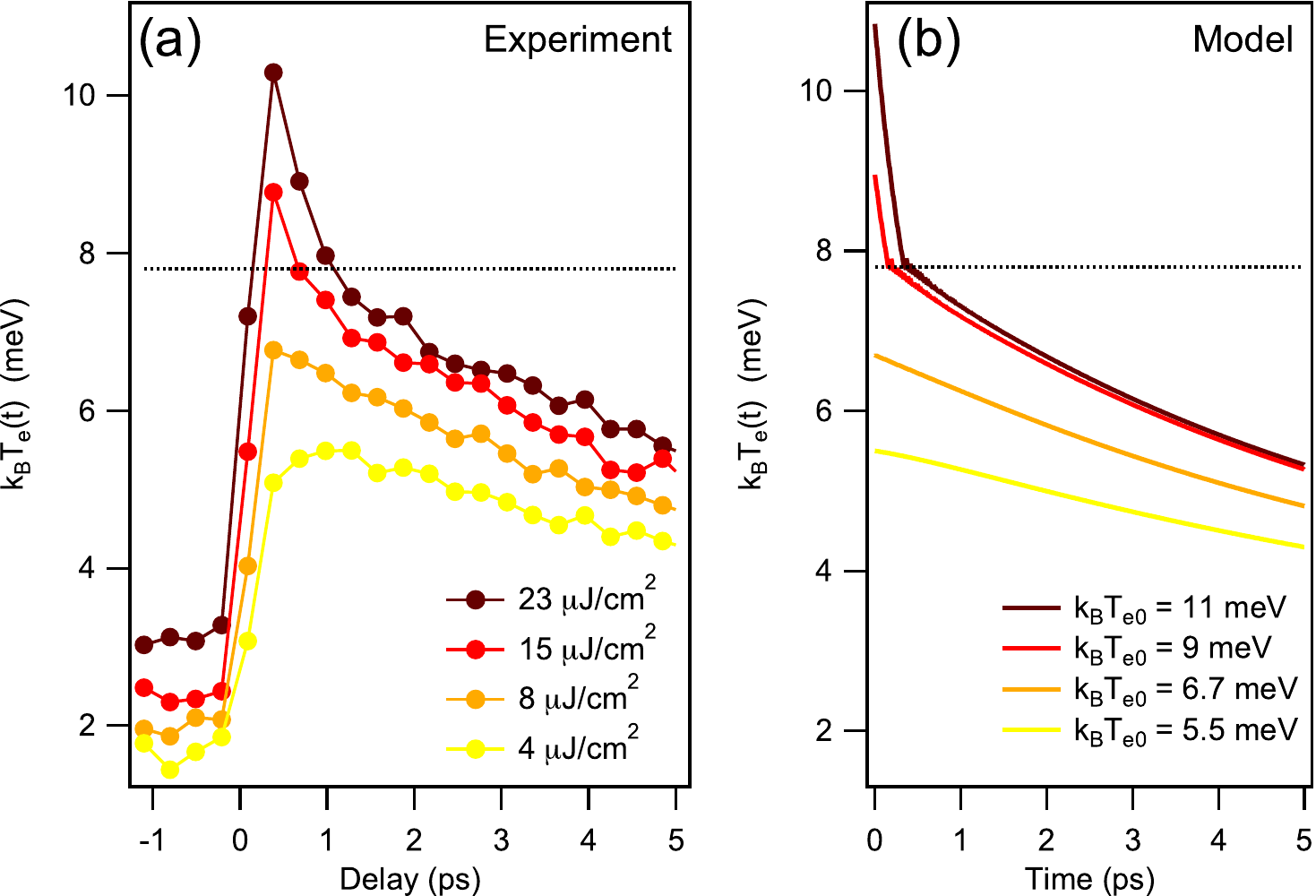}
\caption{\label{expt}Comparison between experimental quasiparticle relaxation dynamics in Bi2212, as measured by time-resolved ARPES, and bosonic relaxation model.
{\bf(a)} Fluence-dependent effective electronic temperature in Bi2212 from an ARPES cut along $\Gamma$-$Y$ at an equilibrium temperature $T=20$ K (see Ref.~\cite{Smallwood15}), obtained by integrating data in $k$ between $k_F-0.08$ $\pi/a$ \AA$^{-1}$ and $k_F+0.08$ $\pi/a$ \AA$^{-1}$ (where $a = 3.83$ \AA) and then fitting to a resolution-convolved Fermi-Dirac distribution function at each delay time \cite{Perfetti07,Smallwood14}. 
{\bf(b)} Theoretical evolution of the effective electronic temperature [defined by Eq.~\refeq{telec}] based on the dynamic-gap quasiparticle relaxation model with a critical temperature $k_BT_c = 7.8$ meV, and with $\alpha^2F(\Omega)$ defined as in Fig.~\ref{rates}(a)\@. The initial electronic distribution corresponds to four different initial electronic temperatures $k_BT_{e0}$. Lattice temperatures $k_BT_b$ were held constant and selected to match the measured electronic temperatures before $t=0$ ps in (a).}
\end{figure}

Figure~\ref{expt} shows a comparison between the model and nodal quasiparticle dynamics in a superconducting sample of Bi2212 near optimal doping ($T_c = 91$ K). The experimental data, which are shown in Fig.~\ref{expt}(a), have been acquired by measuring the pump-induced increase in ARPES intensity above the chemical potential, integrated across a window in energy and momentum along the $\Gamma$-$Y$ direction in $k$ space (see Ref.~\cite{Smallwood15} for further analysis and additional data). Because this momentum cut intersects a superconducting gap node, to a certain extent its temporal dynamics can be viewed as being proportional to the electronic temperature parameter introduced in Section~\ref{sec:qpevolution}\@. Fig.~\ref{expt}(b) shows the evolution of $k_BT_e$ according to the model, for four initial values of $k_BT_e$, two of which are below the critical threshold $k_BT_e = 7.8$ meV and two of which are above it.

In previous work we showed that the fluence threshold for the onset of two-component dynamics in the experimental data is very similar to the threshold at which the superconducting gap closes \cite{Smallwood15}. As shown in the figure, the present model captures the onset of these two-component dynamics quite effectively: When the initial electronic temperature is not sufficiently high to result in a closed gap [yellow and orange theoretical curves in Fig.~\ref{expt}(b)], quasiparticle dynamics evolve smoothly and on the picosecond timescale; When the initial electronic temperature is sufficiently high to result in a closed gap [red and maroon curves in Fig.~\ref{expt}(b)], two-component dynamics emerge, with distinct femtosecond and picosecond timescales.

There are some inconsistencies in the literature as to what the critical fluence value is for the onset of two-component dynamics. The authors of Ref.~\cite{Cortes11} report single-component decay up to a fluence of 32 \uJcm. The authors of Ref.~\cite{Piovera15} report a critical fluence near 40 \uJcm. We have found that an accurate measure of the onset for two-component dynamics can only be obtained after a careful treatment of pump-induced changes in the chemical potential~\cite{Miller15a}, which may resolve this discrepancy.

\section{Conclusions}

In conclusion, we have developed a computationally inexpensive model to explain boson-assisted quasiparticle relaxation dynamics in the presence of both a static and dynamically changing nodal density of states. Although less rigorous than the fully momentum- and energy-dependent Keldysh contour methods, the simplicity of this model makes it useful for developing physical insights into the dynamics of quasiparticles that may be obscured in more complicated approaches. At the same time, the model allows greater flexibility than $N$-temperature models that are adequate for describing the dynamics of a metal, but which---as we have shown here---break down if applied to a system exhibiting a nontrivial density of states. 

Comparisons with time-resolved ARPES experiments reveal that the model captures many of the features of quasiparticle relaxation following an ultrafast near-infrared pump pulse in the cuprate superconductor Bi2212.  The model most prominently captures the transition from femtosecond-scale relaxation dynamics to picosecond-scale relaxation dynamics that is correlated to the opening of the superconducting gap.

Beyond its utility in elucidating superconductor dynamics, the model may be useful in clarifying the dynamics of other types of systems where quasiparticles relax amid nontrivial or gapped densities of states. Recently, for example, a number of groups have begun using time-resolved ARPES to study ultrafast dynamics in graphene \cite{Johannsen13,Gierz13,Ulstrup14}. The model may be particularly useful in studying the dynamics of intrinsically doped graphene, as the density of states at the Dirac point in this material has a similar structure to the density of states of a nodal superconductor.

\appendix

\section{Connections to electronic self-energy}
\label{sec:selfenergy}

It is noted in the main text that the decay rates listed in Eqs.~\refeq{fermigolden}-\refeq{recomb} constitute a different quantity from the imaginary part of the electronic self-energy, specifically the near-equilibrium self-energy values calculated by Kaplan et al.\ \cite{Kaplan76}. The two quantities are still related, however. A lifetime related to the imaginary part of the self-energy $\Sigma''$ can be extracted using a Fermi's golden rule approach through the definition
\begin{equation}
\frac{1}{\tau(\omega)} \equiv -\frac{\partial f_e(\omega)/\partial t}{\Delta f_e(\omega)}, \label{lifetime}
\end{equation}
where the quantity $\Delta f_e(\omega) \equiv f_e(\omega) - f(\omega)$ is the difference between the nonequilibrium distribution function $f_e(\omega)$ and its equilibrium value $f(\omega) \equiv 1/(e^{\omega/k_BT}+1)$, which is the Fermi-Dirac distribution function at equilibrium temperature $T$.

A special case of this formalism is the near-equilibrium situation in the relaxation-time approximation \cite{Grimvall,Ashcroft}, in which $f_e(\omega + \Omega)$ is shifted away from a Fermi-Dirac distribution at temperature $T$ only slightly, and only in the vicinity of $\Omega \approx 0$, such that $n(\Omega)$ and $f_e(\omega \pm \Omega)$ retain their equilibrium values in Eqs.~\refeq{fermigolden2}--\refeq{recomb}, but yet $f_e(\omega)$ is replaced by $f(\omega)+\Delta f_e(\omega)$. Detailed balance requires that $\partial f(\omega)/\partial t = 0$, so we can rewrite Eq.~\refeq{lifetime} as
\begin{align}
\frac{1}{\tau(\omega)} &= -\frac{\partial \Delta f_e(\omega)/\partial t}{\Delta f_e(\omega)} \label{lifetime2} \\[10pt]
&= \frac{\Delta \Gamma_s(\omega)+\Delta \Gamma_r(\omega)}{\Delta f_e(\omega)}, 
\end{align}
where $\Gamma_s(\omega)$ and $\Gamma_r(\omega)$ are defined in analogy to Eqs.~\refeq{scatter} and \refeq{recomb} of the main text as
\begin{align}
\Delta \Gamma_s(\omega) &\equiv \frac{2 \pi}{\hbar} \int_0^\omega d\Omega \, \alpha^2 F(\Omega) \,  D_1(\omega-\Omega) \times \label{neareq1} \\
&\qquad\quad \{ \Delta f_e(\omega)[1-f(\omega-\Omega)][n(\Omega)+1] \nonumber \\
&\qquad\qquad - [-\Delta f_e(\omega)]f(\omega-\Omega)n(\Omega) \}     \nonumber       \\
&\qquad+ \frac{2 \pi}{\hbar} \int_0^\infty d\Omega \, \alpha^2 F(\Omega) \,  D_1(\omega+\Omega) \times \nonumber \\
&\qquad\quad \{ \Delta f_e(\omega)[1-f(\omega+\Omega)]n(\Omega) \nonumber \\
&\qquad\qquad - [-\Delta f_e(\omega)]f(\omega+\Omega)[n(\Omega)+1] \}            \nonumber
\intertext{and}
\Delta \Gamma_r(\omega) &\equiv \frac{2 \pi}{\hbar} \int_\omega^\infty d\Omega \, \alpha^2 F(\Omega) \,  D_1(\omega-\Omega) \times \label{neareq2} \\
&\qquad\quad \{ \Delta f_e(\omega)[1-f(\omega-\Omega)][n(\Omega)+1] \nonumber \\
&\qquad\qquad - [-\Delta f_e(\omega)]f(\omega-\Omega)n(\Omega) \}.            \nonumber
\end{align}
Eq.~\refeq{lifetime2} can be simplified by eliminating the terms from Eqs.~\refeq{neareq1} and \refeq{neareq2} that additively cancel, and making use of the identity $1-f(\omega') = f(-\omega')$, to arrive at
\begin{align}
&\frac{1}{\tau(\omega)} = \label{neareq} \\
&\frac{2 \pi}{\hbar} \int_0^\omega d\Omega \, \alpha^2 F(\Omega) \,  D_1(\omega-\Omega) \{ f(\Omega - \omega)+n(\Omega) \}            \nonumber \\
&+\frac{2 \pi}{\hbar} \int_0^\infty d\Omega \, \alpha^2 F(\Omega) \, D_1(\omega+\Omega) \{  f(\omega + \Omega)+n(\Omega) \} \nonumber \\
&+\frac{2 \pi}{\hbar} \int_\omega^\infty d\Omega \, \alpha^2 F(\Omega) \, D_1(\omega-\Omega) \{ f(\Omega-\omega)+n(\Omega) \}. \nonumber
\end{align}
The first term of Eq.~\refeq{neareq} corresponds to scattering between states at $\omega$ and states at lower energy, the second term corresponds to scattering between states at $\omega$ and states at higher energy, and the third term corresponds to recombination/pair breaking interactions.
A similar form of this derivation can be found in Ref.~\cite{Grimvall}.

We note here, as well as in the main text, that in the present formulation the role of the superconducting condensate has been ignored apart from its impact on the density of states. However, one can extend the expression to a more accurate form for superconductors without significant difficulty, by multiplying in appropriate coherence factors before each of the terms in Eq.~\refeq{neareq}, and by dividing out an overall renormalization factor $Z_1(0)$. For an $s$-wave gap, this leads to 
\begin{align}
\frac{1}{\tau(\omega)} = &\frac{2 \pi}{\hbar \, Z_1(\omega)} \times \label{neareqswave} \\
\left[ \vphantom{\int_0^\omega} \right. &\int_0^\omega d\Omega \, \alpha^2 F(\Omega) \left\{ 1-\frac{\Delta^2}{\omega(\omega-\Omega)}\right\} \nonumber \\
&\qquad  D_1(\omega-\Omega) \{ f(\Omega - \omega)+n(\Omega) \}            \nonumber \\
+&\int_0^\infty d\Omega \, \alpha^2 F(\Omega) \left\{ 1-\frac{\Delta^2}{\omega(\omega-\Omega)}\right\} \nonumber \\
&\qquad D_1(\omega+\Omega) \{  f(\omega + \Omega)+n(\Omega) \} \nonumber \\
+&\int_\omega^\infty d\Omega \, \alpha^2 F(\Omega) \left\{ 1-\frac{\Delta^2}{\omega(\omega+\Omega)}\right\} \nonumber \\
&\qquad D_1(\omega-\Omega) \{ f(\Omega-\omega)+n(\Omega) \} \left. \vphantom{\int_0^\omega} \right],\nonumber
\end{align}
which replicates Eq.~6 from Kaplan et al.\ \cite{Kaplan76}, where quasiparticle lifetimes near equilibrium in conventional superconductors are derived using a more rigorous Green's-function approach. The case of a $d$-wave superconductor is more complicated because the coherence factors are momentum-dependent, but follows the same basic logic.

\section{Rothwarf-Taylor model}
\label{sec:rtmodel}

Further simplifications of Eq.~\refeq{fermigolden2} can be used to establish a direct connection to the Rothwarf-Taylor model of quasiparticle relaxation \cite{Rothwarf67}. One can define an energy-integrated quasiparticle population $p$ according to
\begin{equation}
p \equiv 2\int_0^\infty d\omega D(\omega) f_e(\omega).
\end{equation}
Analyzing the typical features of $D(\omega)$ and $f(\omega)$, the bulk of the quasiparticle population $p$ will often be localized near the gap edge at energy $\Delta$. Thus, we are in certain cases justified in approximating $D(\omega)$ by the delta-function expression 
\begin{equation}
D(\omega) \approx D_\Delta \delta(\omega-\Delta)+D_\Delta\delta(\omega+\Delta), \label{dta}
\end{equation}
where $D_\Delta \equiv p(t_\text{ref})/[2 f_e(\Delta,t_\text{ref})]$ is a constant and $t_\text{ref}$ is an arbitrarily selected reference time.
One can then multiply Eq.~\refeq{fermigolden2} by $D(\omega)$, substitute Eqs.~\refeq{dta} and \refeq{recomb} into the result, and integrate over $\omega$ to obtain
\begin{align}
\dot{p} &= -\frac{\pi \alpha^2 F(2\Delta)}{\hbar D_0} \times 
\Big\{p^2 [n(2\Delta,T_b)+1] \label{recomb2}\\
&\qquad \quad - (4 D_\Delta^2 - 4 D_\Delta p - p^2) n(2\Delta,T_b)\Big\}. \nonumber
\intertext{Finally, if we define a boson population $N \equiv F(2 \Delta) n(2\Delta)$ and consider the limits $D_{\Delta} \gg p$ and $n(2\Delta,T_b) \ll 1$, we arrive at the expression}
\dot{p} &= -\frac{ \pi \alpha^2 F(2\Delta)}{\hbar D_0} \left[p^2 - \frac{4 D_\Delta^2}{F(2\Delta)} N \right].
\end{align}
After a trivial change of variables, this can be rewritten as $\dot{p} = -Rp^2 + \gamma N$, which is the first equation of the Rothwarf-Taylor model.

This derivation highlights some of the important physical origins of the Rothwarf-Taylor model's recombination coefficients $R$ and $\gamma$. It also demonstrates some failures of the model. For example, scattering processes are ignored in the Rothwarf-Taylor model, and the Rothwarf-Taylor model is only truly appropriate at small $N$. As $N$ becomes increasingly large, stimulated emission processes may become relevant [as encapsulated, for example, by Eq.~\refeq{recomb2}].

\section{Incremental dependence of $P_1(\omega,t)$ on $\Delta(\omega,t)$}
\label{sec:gapopen}

Equation \refeq{openingdwave} is derived as a $d$-wave-gap extension of a condition under an $s$-wave gap requiring quasiparticle population to remain conserved for a given value of $k$ as the gap opens; that is, it is derived under the assumption that states transform adiabatically between time steps. For an $s$-wave gap in BCS theory, the opening of the superconducting gap amounts to a band structure that is modified relative to its normal-state parent structure according to
\begin{equation}
E_k^2 = \xi_k^2 + \Delta_k^2, \label{bcs}
\end{equation}
where $E_k$ is the superconducting state band energy, $\xi_k$ is the normal-state band energy, and $\Delta_k$ is the gap parameter \cite{Tinkham}. Because of this relationship, mappings between $E_k$ values of different gap parameters $\Delta_k$ are bijective, if it is also understood that the two values of $E_k$ carry the same sign. In consequence, the requirement that quasiparticle number be conserved for an incrementally opening gap amounts to a requirement that 
\begin{equation}
f_e(k,E_{k,i+1},t_{i+1}) = f_e(k,E_{k,i},t_{i}), \label{swavebcs}
\end{equation}
where the energy arguments $E_{k,i}$ and $E_{k,i+1}$ are related to each other through the BCS relationship \refeq{bcs} according to
\begin{equation}
E_{k,i}^2 = E_{k,i+1}^2 - (\Delta_{k,i+1}^2 - \Delta_{k,i}^2).
\end{equation} 
It is therefore possible to express $f_e(k,\omega_k,t_{i+1})$ exclusively in terms of the time-dependent gap magnitudes $\Delta_{k,i}$ and $\Delta_{k,i+1}$ and the information about $f_e$ available at a proximate time $t_i$:
\begin{align}
f_e&(k,\omega_k,t_{k,i+1}) \label{swavebcs2} \\
&= f_e\left(k,\sqrt{\omega_k^2 - [\Delta_{k,i+1}^2 - \Delta_{k,i}^2]},t_i\right). \nonumber
\end{align} 

To adapt this for a $d$-wave gap, we approximate $\Delta_{k,i}$ as an angular function, where momentum dependence is captured by the Fermi surface angle $\phi_k$ such that $\Delta_{k,i} \to \Delta_{\phi,i}$, and where $\Delta_{\phi,i}$ is defined according to the relationships outlined in Section \ref{sec:decayrates}:
\begin{equation}
\Delta_{\phi,i} = \left\{ 
\begin{array}{ll}
\frac{4 \Delta_i}{\pi} \left(\phi_k - \frac{\pi}{4}\right) & \textrm{first BZ quadrant,} \\[5pt]
-\Delta_{\phi + \pi/2,i} & \textrm{other quadrants.}
\end{array}
\right.
\end{equation}
To simplify notation, we perform a variable substitution $u \equiv 4 \Delta_{i+1}\times(\phi_k + \pi/4)/\pi$ such that $u = \Delta_{k,i+1}$, which leads to the expression
\begin{align}
f_e&(u,\omega,t_{i+1}) \label{swavebcs3} \\
&= f_e\left(u,\sqrt{\omega^2 - [1 - (\Delta_{i}/\Delta_{i+1})^2]u^2},t_i\right). \nonumber
\end{align} 
We then multiply by the $u$-dependent density of states $\omega/\sqrt{\omega^2-u^2}$ to convert $f_e$ into a momentum-dependent quasiparticle population, 
\begin{align}
D_{1}&(u,\omega,t_{i+1})f_e(u,\omega,t_{i+1}) \label{dwavebcs}\\[5 pt]
&= \frac{\omega}{\sqrt{\omega^2-u^2}} f_{e}\left(u,\sqrt{\omega^2 - [1-(\Delta_{i}/\Delta_{i+1})^2]u^2} ,t_i\right).\nonumber 
\end{align}
Finally, we assume that $f_e$ was momentum-independent at time $t_i$, and average out the momentum dependence of $D_1(u,\omega,t_{i+1})f_e(u,\omega,t_{i+1})$ by performing a normalized integral of Eq.~\refeq{dwavebcs} with respect to $u$ between 0 and $\min(\omega,\Delta_{i+1})$, resulting in Eq.~\refeq{openingdwave}.

\begin{acknowledgments}
We thank A.~F.\ Kemper, J.\ Orenstein, D.-H.\ Lee, J.~P.\ Hinton, and Z.\ Tao for useful discussions, and H. Eisaki for providing material samples.
This work was supported as part of the Ultrafast Materials Program at Lawrence Berkeley National Laboratory, funded by the U.S.\ Department of Energy, Office of Science, Office of Basic Energy Sciences, Materials Sciences and Engineering Division, under Contract No.\ DE-AC02-05CH11231\@. C.L.S. acknowledges partial support from an NRC Research Associateship award at NIST.
\end{acknowledgments}

%

\end{document}